\pdfoutput=1
\documentclass[aps,prd,preprint,superscriptaddress,tightenlines,nofootinbib]{revtex4}

\long\def\symbolfootnote[#1]#2{\begingroup%
\def\thefootnote{\fnsymbol{footnote}}\footnote[#1]{#2}\endgroup}

\usepackage{amsfonts,amsmath,amsthm,amssymb}
\usepackage{mathrsfs}
\usepackage{bm}
\usepackage{color}
\usepackage{epsfig}
\newcommand{\PRE}[1]{{#1}}   

\DeclareMathAlphabet{\mathpzc}{OT1}{pzc}{m}{it}

\newcommand{\beq}{\begin{equation}}
\newcommand{\eeq}{\end{equation}}
\newcommand{\bea}{\begin{flushleft} \begin{eqnarray}}
\newcommand{\eea}{\end{eqnarray}\end{flushleft}}

\newcommand{\postscript}[2]{\setlength{\epsfxsize}{#2\hsize}
   \centerline{\epsfbox{#1}}}
\newcommand{\comment}[1]{}

\newcommand{\el}[1]{\label{#1}}

\newcommand{\df}[1]{\textbf{#1}}

\newcommand{\ci}[1]{}

\newcommand{\ke}{\rangle}
\newcommand{\br}{\langle}
\newcommand{\lb}{\left(}
\newcommand{\rb}{\right)}

\newcommand{\ba}{\begin{eqnarray}}
\newcommand{\ea}{\end{eqnarray}}
\newcommand{\be}{\begin{equation}}
\newcommand{\ee}{\end{equation}}
\newcommand{\bay}[1]{\left(\begin{array}{#1}}
\newcommand{\eay}{\end{array}\right)}

\newcommand{\zt}[1]{\rm{#1}}

\def\xa{{\alpha}}

\def\xg{{\gamma}}

\def\xL{{\Lambda}}

\def\xs{{\sigma}}

\def\xt{{\theta}}

\def\CQ{{\cal Q}}

\usepackage[usenames,dvipsnames]{xcolor}
\definecolor{orange}{cmyk}{0,0.5,1,0}
\definecolor{rossoCP3}{cmyk}{0,.88,.77,.40}
\definecolor{graa}{rgb}{0.8,0.8,0.8}
\definecolor{blaa}{rgb}{0.2,0.2,0.6}

\begin{document}

\title{\PRE{\vspace*{0.9in}} \color{rossoCP3}{Constraints on 750~GeV
 colorless  $\mathcal Q$-onia from running couplings 
}}

\author{{\bf Luis A. Anchordoqui}}

\affiliation{Department of Physics and Astronomy,\\  Lehman College, City University of
  New York, NY 10468, USA
\PRE{\vspace*{.05in}}
}

\affiliation{Department of Physics,\\
 Graduate Center, City University
  of New York,  NY 10016, USA
\PRE{\vspace*{.05in}}
}

\affiliation{Department of Astrophysics,\\
 American Museum of Natural History, NY
 10024, USA
\PRE{\vspace*{.05in}}
}

\author{{\bf Haim \nolinebreak Goldberg}}
\affiliation{Department of Physics,\\
Northeastern University, Boston, MA 02115, USA
\PRE{\vspace*{.05in}}
}

\author{{\bf Xing Huang}}
\affiliation{Department of Physics, \\
National Taiwan Normal University, Taipei, 116, Taiwan
\PRE{\vspace*{.05in}}
}

\PRE{\vspace*{.15in}}

\begin{abstract}\vskip 2mm
  \noindent We present yet another composite model explaining the
  relatively broad peak in the diphoton invariant mass distribution
  around 750 GeV recently observed at the LHC experiments. We consider
  the excess originates in bound states of vector-like fermions $\cal
  Q$ transforming under the electroweak group (but not QCD) of the
  standard model and which are also charged under a new $SU(N_{\rm
    TC})$ QCD-like confining force.  Since the new uncolored fields
  transform as $SU(2)$ doublets they can acquire a mass through Yukawa
  interactions with the electroweak Higgs doublet, as quarks and
  leptons. We parametrize the ${\mathcal Q}$-onium bound state using
  the Coulomb approximation and give a numerical fit to the diphoton
  excess consistent with no strong coupling effects up to about
  30~TeV, which is the energy scale for next generation colliders.
  The range of allowed hypercharge $Y_{\mathcal Q}$ is given by $2.26
  \leq Y_{\mathcal Q} \leq 2.53$ for $N_{\rm TC} =2$ and $2.17 \leq
  Y_{\mathcal Q} \leq 2.31$ for $N_{\rm TC} = 3$. The smoking gun for
  the model would be the discovery of $\mathcal{Q}$-onia decaying into
 Higgs and $Z$-boson.

\end{abstract}

\maketitle

\section{Introduction}

In December 2015, ATLAS~\cite{ATLAS} and CMS~\cite{CMS:2015dxe}
famously announced the observation of a peak in the diphoton mass
distribution around 750~GeV, using (respectively) $3.2~{\rm fb}^{-1}$
and $2.6~{\rm fb}^{-1}$ of data recorded at a center-of-mass energy
$\sqrt{s} = 13~{\rm TeV}$.  The diphoton excesses could be interpreted
as the decay products of a new massive particle $X$, with spin 0, 2,
or higher~\cite{Landau:1948kw}.  Assuming a narrow width approximation
ATLAS gives a local significance of $3.6\sigma$, or else a global
significance of $2.0\sigma$ when the look-elsewhere-effect in the mass
range $M_X/{\rm GeV} \in [200 - 2000]$ is accounted for. Signal-plus-background fits were also implemented for a broad
signal component with a large decay width. The largest 
deviation from the background-only hypothesis corresponds to $M_X \sim
750~{\rm GeV}$ with a total width \mbox{$\Gamma_{\rm total} \sim
  45~{\rm GeV}$.} The local and global significances evaluated for the
broad resonance fit are roughly 0.3 higher than that for the fit using
the narrow width approximation, corresponding to $3.9\sigma$ and
$2.3\sigma$, respectively. The CMS data gives a local significance of
$2.6\sigma$ and a global significance smaller than
$1.2\sigma$. Fitting the LHC data at $\sqrt{s} = 13~{\rm TeV}$ to a
resonant structure leads to a cross section times branching fraction of
\begin{equation}
\sigma_{\rm LHC13} (pp \to X +
{\rm 
anything}) \times {\cal B} (X \to \gamma \gamma) \sim
\left\{\begin{array}{cl} 
(10 \pm 3)~{\rm fb} &~~~ {\rm ATLAS} \\
\phantom{0}(6 \pm 3)~{\rm fb} &~~~ {\rm CMS} \end{array}
\right. \,,
\end{equation}
at $1\sigma$~\cite{Franceschini:2015kwy}. However, the LHC data at
$\sqrt{s} = 8~{\rm TeV}$ show no excess over standard model processes
(SM)~\cite{Aad:2015mna,Khachatryan:2015qba}, placing a critical upper
bound on the cross section times branching fraction: $\sigma_{\rm LHC8}
(pp \to X + {\rm anything}) \times {\cal B} (X \to \gamma
\gamma) < 2.00~{\rm fb}$ at 95\% CL~\cite{Khachatryan:2015qba}.
Altogether, the LHC8 data would be compatible with the LHC13 data within
$\sim 2 \sigma$, if the diphoton cross section grows by more than
about a factor of 3 or 3.5.

Quite recently, ATLAS and CMS updated their diphoton resonance
searches~\cite{Delmastro,Musella,CMS:2016owr}. ATLAS reanalyzed the
$3.2~{\rm fb}^{-1}$ of data, targeting separately spin-0 and spin-2
resonances. For spin-0, the most significant deviation from the
background-only hypothesis corresponds to $M_X \sim 750~{\rm GeV}$ and
$\Gamma_{\rm total} \sim 45~{\rm GeV}$. The local significance is now
increased to $3.9\sigma$ but the global significance remains at the $2
\sigma$ level. For the spin-2 resonance, both the local and global
significances are reduced down to $3.6\sigma$ and $1.8\sigma$,
respectively. The new CMS analysis includes additional data (recorded
in 2015 while the magnet was not operated) for a total of 3.3~${\rm
  fb}^{-1}$. The largest excess is observed for $M_X = 760~{\rm GeV}$
and $\Gamma_{\rm total} \approx 11~{\rm GeV}$,  and has a local
significance of $2.8\sigma$ for spin-0 and $2.9\sigma$ spin-2
hypothesis. After taking into account the effect of searching for
several signal hypotheses, the significance of the excess is reduced
to $< 1 \sigma$. CMS also communicated a combined search with data
recorded at $\sqrt{s} = 13~{\rm TeV}$ and $\sqrt{s} = 8~{\rm
  TeV}$. For the combined analysis, the largest excess is observed at
$M_X = 750~{\rm GeV}$ and $\Gamma_{\rm total} = 0.1~{\rm GeV}$. The
local and global significances are $\approx 3.4\sigma$ and
$1.6\sigma$, respectively.

Among many possible interpretations~\cite{Staub:2016dxq}, one of the
most attractive type of models is that of pairs of new heavy fermions
which produce resonant signals of their near-threshold bound
states. It seems natural to classify the various models in accordance
with the fermion properties. The simplest explanation for the
structure underlying the bump in the diphoton spectrum could be that
of a QCD bound state of a heavy vector-like quark $Q$, with a mass around
375~GeV~\cite{Barger:1987xg,Kats:2012ym,Luo:2015yio,Zhang:2016xei,Han:2016pab,Kats:2016kuz,Hamaguchi:2016umx}.
For this model, the signal rate can be explained by a scalar $Q \bar
Q$ bound state $\eta_Q$, with $Q$ transforming as $(3, 1, -4/3)$ under
the $SU(3) \otimes SU(2) \otimes U(1)_Y$ gauge groups of the
SM~\cite{Han:2016pab,Kats:2016kuz}.  Actually, in analogy to $\eta_c$
and $J/\psi$, the QCD bound states would contain both scalars and
vectors.  We may note in passing that the vector $J/\psi$-like state
will not only decay into $\gamma \gamma$ but also into dilepton
topologies. The latter is severely constrained by
ATLAS~\cite{Aad:2014cka} and CMS~\cite{Khachatryan:2014fba} direct
searches; however, since the $J/\psi$-like state can only be produced
through $q \bar q$ annihilation, its production cross section at the
LHC is significantly smaller than that of the $Q \bar Q$ scalar state.

A somewhat related model interprets the excess of diphoton events as
arising also from the decay of a $\mathpzc Q$-onium state $\eta
_{\mathpzc Q}$, but with the colored fermions ${\mathpzc Q}$ bound by
a hidden confining
$SU(N)$~\cite{Harigaya:2015ezk,Nakai:2015ptz,Curtin:2015jcv,Bian:2015kjt,Craig:2015lra,Redi:2016kip,Kamenik:2016izk,Ko:2016sht,Bai:2016czm}. One
particularly interesting possibility to describe the hidden gauge
dynamics is given by $SU(N_{\rm TC})$, with $N_{\rm TF}$
techni-flavors. This is because the dynamics of this theory is
renowned from QCD and can be also understood in the large $N_{\rm TC}$
limit: the gauge theory is asymptotically free provided it satisfies
the familiar bound on the number of techini-flavors and confines at a
scale $\Lambda_{\rm TC}$. To avoid the strict constraints common to
old techni-color models the heavy-fermions must be in a vectorial
representation of the SM and in the fundamental $\bm{N}_{\rm TC}$ of
$SU(N_{\rm TC})$; namely
\begin{equation}
{\mathpzc Q} = \sum_{i =1}^{N_s} {\mathpzc Q}_{\  i}, \quad {\rm with}
\quad
{\mathpzc Q}_{\ i} =
(\bm{N}_{\rm TC}, \bm{R}_i) \oplus (\bm{\bar{ N}}_{\rm TC}, \bm{\bar{R}}_i) \,,
\label{vec-rep}
\end{equation}
where $\bm{R}_i$ denotes a generic SM representation and $N_s$ is the
number of species with mass below the confinement
scale~\cite{Redi:2016kip}. Yet a third dynamics is possible if
the heavy fermions $\mathcal{Q}$ are colorless particles, bound by a
confining $SU(N_{\rm TC})$~\cite{Iwamoto:2016ral}. 

At this point it is important to stress two fundamental differences
between these three models. Firstly, while $\eta_Q$ and
$\eta_{\mathpzc Q}$ could be produced either via gluon or photon
fusion in LHC collisions, the production of $\eta_\mathcal{Q}$ could
only proceed via photon fusion. Secondly, if the heavy particles
belong to the $SU(2)$ doublet and feel the strong $SU(3)$ color 
interactions, then the 125 GeV~Higgs signal strength in the $gg \to H
\to \gamma \gamma$ channel would be significantly modified.
ATLAS~\cite{Aad:2015gba} and CMS~\cite{Khachatryan:2014jba} data would
then place severe constraints on model parameters unless the coupling
of $Q$ or ${\mathpzc Q}$ to the Higgs is small.  However,
$\mathcal Q$ transforms as an $SU(3)$ singlet and so it does not
contribute to $gg \to H \to \gamma \gamma$. Therefore, the $\mathcal Q$'s could
get mass like quarks and leptons, as $SU(2)$ doublets and singlets can form a
mass term after being Higgsed. This is precisely what motivates our
study.

In this paper we investigate the phenomenology of $\mathcal Q$-onia
states of a hidden $SU(N_{\rm TC})$.  We combine the requirements to
reproduce the LHC diphoton signal with those arising from the
renormalization group (RG) equations to constrain the parameter space
of $\mathcal{Q}$ fermions transforming under the electroweak group
(but not QCD) of the SM. For related RG studies in which the 750~GeV
excitation is not a bound state,
see~\cite{Gu:2015lxj,Son:2015vfl,Goertz:2015nkp,Dev:2015vjd,Gross:2016ioi}. Before
proceeding we note that the LHC phenomenology of $\mathcal Q$-onia
states, with constituents that carry only hypercharge, but no SM
$SU(3)$ or $SU(2)$ quantum numbers has been presented
in~\cite{Iwamoto:2016ral}. Though at first sight our study seems quite
similar to the analysis in~\cite{Iwamoto:2016ral}, it differs in a
fundamental aspect: herein we consider colorless fermions that
transform as $SU(2)$ doublets and thus the heavy $\mathcal Q$
particles can obtain a mass, as all other SM fermions, through the
Higgs mechanism. The layout of the paper is as follows. In
Sec.~\ref{secII} we introduce the Lagrangian of the model with a new
confining gauge interaction and derive the RG equations. In
Sec.~\ref{secIII} we discuss the LHC phenomenology of our set up. We
parametrize the $\mathcal{Q}$-onium bound state using the Coulomb
approximation~\cite{Kats:2012ym} and give a numerical fit to the
diphoton excess consistent with the running couplings. We request no
strong coupling effects up to about 30~TeV, which is the energy scale
for next generation colliders~\cite{Gershtein:2013iqa}. In
Sec.~\ref{secIV} we verify consistency with early universe
cosmology. Our conclusions are collected in Sec.~\ref{secV}.

\section{Lagrangian and Renormalization Group equations}
\label{secII}

To develop our program in the simplest way, we will work within the construct of a minimal model in which we consider one generation of heavy colorless
fermions, which contains two $SU(2)$ doublets (one left- and one
right-handed to make the representation vectorial, i.e. symmetric
under parity) and four chiral fermions that transform as $SU(2)$
singlets (two of them are left-handed and the other two
right-handed). We label the four flavors as club,
diamond, heart, and spade, and so we named the $\mathcal Q$ particles
``qards'' accordingly.\footnote{Note that qards are quirks
  transforming under the electroweak group (but not QCD) of the
  SM~\cite{Kang:2008ea,Kribs:2009fy,Martin:2010kk,Harnik:2011mv,Fok:2011yc}.}
The Lagrangian for the technicolor qards is
given by
\begin{eqnarray} \nonumber
\mathscr{L}_{\mathcal Q} &=&  \Big( i\overline {\bm{\mathcal Q}_{L}} \gamma
_{\mu}D^{\mu} \bm{\mathcal{Q}}_{L} + i\overline {\mathcal Q^{_\clubsuit}_{R}}
                        \gamma _{\mu} D^{\mu}  \mathcal Q^{_\clubsuit}_R
                        +i\overline {\mathcal Q^{_\diamondsuit}_R} \gamma
                        _{\mu}D^{\mu} \mathcal Q^{_\diamondsuit}_R +\\
                          &+ &  i\overline {\bm{\CQ_{R}}} \gamma
                          _{\mu}D^{\mu} \bm{\CQ}_{R} + i\overline 
                                                  {\mathcal Q^{_\heartsuit} _{L}}
                        \gamma _{\mu}D^{\mu} \mathcal Q^{_\heartsuit}_L
                        +i\overline {\mathcal Q^{_\spadesuit}_L} \gamma
                        _{\mu}D^{\mu} \mathcal Q^{_\spadesuit}_L \Big)  +\zt{h.c.}\, ,
\end{eqnarray}
where the covariant derivative for the $SU(2)$ doublets $\bm{\CQ}_{L,R}$, 
\begin{equation}
\bm{\CQ}_L =  \bay{c}
 \mathcal Q^{_\clubsuit}_L \\ \mathcal Q^{_\diamondsuit}_L \eay \quad
 {\rm and} \quad \bm{\CQ}_R =  \bay{c}
 \mathcal Q^{_\heartsuit}_R \\ \mathcal Q^{_\spadesuit}_R \eay \,,
\end{equation}
reads 
\be
{D}_\mu = \partial_\mu  - i g_2 \tau^a A^a_\mu - i g_{\zt{TC}} \tilde T^a
\tilde G^a_\mu- i g_Y Y_\CQ B_\mu\,,
\ee
while those for SU(2) singlets $\mathcal Q^{_\heartsuit}_{L}$, $\mathcal
Q^{_\spadesuit}_{L}$, $\mathcal Q^{_\clubsuit}_{R}$, $\mathcal Q^{_\diamondsuit}_{R}$ are (plus
sign for $\mathcal Q^{_\clubsuit}_R$, $\mathcal Q^{_\heartsuit}_L$)
\be
{D}_\mu = \partial_\mu   - i g_{\zt{TC}} \tilde T^a
\tilde G^a_\mu- i g_Y \left(Y_\CQ \pm \tfrac 1 2 \right)B_\mu\,.
\ee
The Yukawa interactions that provide the masses to the qards are:
\be
\mathscr{L}_{\cal Y} = -{\cal Y}_{_{\! \diamondsuit}}\overline {\bm{\CQ}_{L}} \cdot H \mathcal Q^{_\diamondsuit}_{R} 
                 - {\cal Y}_{_{\! \clubsuit}}\overline {\bm{\CQ}_{L}} \cdot
                 \widetilde H \mathcal Q^{_\clubsuit}_{R} 
                 -{\cal Y}_{_{\!\spadesuit}}\overline {\mathcal Q^{_\spadesuit}_{L}} \bm{\CQ}_{R}\cdot H 
                 -{\cal Y}_{_{\! \heartsuit}}\overline {\mathcal Q^{_\heartsuit}_{L}} \bm{\CQ}_{R}\cdot \widetilde H +\zt{h.c.}\label{L_Yukawa}
\ee
where $\widetilde H=i\sigma^2 H^*$. The couplings of the technicolor qards
with the Higgs are the same as those with SM fermions and hence they get masses
in the same way. Substituting in (\ref{L_Yukawa}) the Higgs vacuum expectation value, 
\begin{equation}
\br H \ke = \frac 1 {\sqrt 2} \bay{c}
0 \\ v \eay , 
\end{equation}
we have the mass terms
\be
\mathscr{L}_m = -\frac 1 {\sqrt 2} {\cal Y}_{_{\!\diamondsuit}}\overline {\mathcal
  Q^{_\diamondsuit}_{L}} \mathcal Q^{_\diamondsuit}_{R}v 
                 -\frac 1 {\sqrt 2} {\cal Y}_{_{\! \clubsuit}}\overline {\mathcal
                   Q^{_\clubsuit}_{L}} \mathcal Q^{_\clubsuit}_{R}v 
                 -\frac 1 {\sqrt 2} {\cal Y}_{_ {\! \spadesuit}}\overline {\mathcal Q^{_\spadesuit}_{L}}
               \mathcal  Q^{_\spadesuit}_{R}v 
                 -\frac 1 {\sqrt 2} {\cal Y}_{_ {\! \heartsuit}}\overline {\mathcal Q^{_\heartsuit}_{L}}
                 \mathcal Q^{_\heartsuit}_{R}v +\zt{h.c.}\label{L_Yukawa_m}
\ee

The RG equation for the coupling of an
$SU(N_c)$ gauge field with $N_f$ fundamental matter reads
\begin{equation}
\frac {d }{d t}g(t)=-\frac{g^3 }{(4 \pi )^2}\left(\frac{11
    N_c}{3}-\frac{2 N_f}{3}\right)\,.
\end{equation}
Throughout we take both the left and right-handed technicolor $SU(N_{\zt{TC}})$ qards to be in the representation of 
\be \el{t-rep}(\bm{ N}_{\zt{TC}},\df 1, \df 2 \oplus \df 1 \oplus \df 1)_Y \ee 
under $SU(N_{\zt{TC}}) \otimes SU(3) \otimes SU(2)  \times U(1)_Y$,
satisfying a relation akin to (\ref{vec-rep}), 
\begin{equation}
{\mathcal Q} =
(\bm{N}_{\rm TC}, \bm{1},\bm{2} \oplus \bm{1} \oplus \bm{1} )_Y \oplus (\bm{\bar{N}}_{\rm TC}, \bm{1},
\bm{2} \oplus \bm{1} \oplus \bm{1})_{-Y} \, .
\end{equation}
Thus, the RG equation become
\be
\frac{d}{d t} g_{\zt{TC}} = \frac{g_{\rm TC}^3}{16\pi ^2}\left[ -\frac {11} 3 N_{\zt{TC}}+\frac{2}{3} \times
4 \right] = \frac{g_{\rm TC}^3}{16\pi ^2}\left( -\frac {11} 3 N_{\zt{TC}}+\frac
8 3\right)\, ,
\ee
For the $SU(3)$ and $SU(2)$ gauge couplings $g_3$ and $g_2$, the RG
equations are
\begin{equation}
\frac{d}{dt}g_3 = \frac{g_3^3}{16\pi ^2}\left[ -11+\frac{2}{3}(2 \times
3)\right] = \frac{g_3^3}{16\pi ^2}\left( -7\right)\, ,\el{rg-g3}
\end{equation}
and 
\begin{equation}
\frac{d}{dt}g_2 = \frac{g_2^3}{16\pi ^2}\left[ -\frac{22}{3}+\frac{2}{3}
  \left(\frac
3 2 \times 3 + \frac 1 2 \times 3 +  N_{\zt{TC}} \right)+\frac{1}{6}\right] =
\frac{g_2^3}{16\pi ^2}\left(-\frac{19}{6}+\frac 2 3 N_{\zt{TC}}\right)
\, .\el{rg-g2} 
\end{equation}
The RG running of the hypercharge coupling can be worked out in a
similar fashion, 
\be \frac{d}{dt}g_Y = \frac{1}{16\pi
  ^2}\left\{\frac{2}{3} \sum _f Y_f^2 + \frac{1}{3}\sum _s Y_s^2 \right\}
   g_Y^3 \, , 
\label{rgy}
\ee where the sum is over the hypercharges of chiral
fermions $f$ and complex scalars $s$. For the case at hand (\ref{rgy}) can be
rewritten as
\begin{equation}
\frac{dg_Y}{dt} = \frac{g_Y^3}{16 \pi^2}
 \left\{ \frac{2}{3}
\times 2 \times \left[2N_{\zt{TC}} Y_{\mathcal
    Q}^2+N_{\zt{TC}} \left(Y_{\mathcal
      Q}-\frac{1}{2}\right)^2+N_{\zt{TC}}
  \left(Y_{\mathcal{Q}}+\frac{1}{2}\right)^2\right]+\frac {41}
6\right\} \!,  
\end{equation} 
where $Y_{\mathcal Q}$ is the hypercharge of the technicolor qards
in the $SU(2)$ doublet. Notice that the qards in the doublet of $SU(2)$
have charge $Y_{\mathcal Q}$ while those in the singlet have charges
$Y_{\mathcal Q} \pm
\frac 1 2$ in order to allow Yukawa coupling. The extra factor of $2$
follows from the fact that we have doublets for both left- and
right-handed qards.

We denote the Yukawa coupling of the qard with hypercharge
$Y_{\mathcal Q}\pm\frac 1 2$ as ${\cal Y}_\pm$. The RG equation for
${\cal Y}_\pm$ is given by 
\begin{equation}\label{RGE_yuk_top}
\frac{d}{d t} {\cal Y}_\pm = \frac{{\cal Y}_\pm}{16\pi
  ^2}\left[\frac{2 N_{\zt{TC}}+3}{2} {\cal Y}_\pm^2+\frac 3 2
{\cal Y}_\mp^2 -\frac {3(N_{\zt{TC}}^2-1)} {N_{\zt{TC}}} g_{\zt{TC}}^2-
\frac{9}{4}g_2^2-\lb\frac{3}{4} \pm 3 Y_Q +6 Y_Q^2 \rb g_Y^2 \right]\, .
\end{equation}
The $SU(2)$ quantum number is the same
as in SM and hence so is the contribution from $SU(2)$ gauge fields. The only
difference is in the hypercharges. It can be easily checked that for
$Y_{\mathcal Q} =  1 /6$, the $g_Y$ contribution ($3/4+3 Y_{\mathcal Q} +6
Y_{\mathcal Q}^2$) gives $17/12$, which is the value for the SM.

\section{Playing qards at the LHC}
\label{secIII}

For simplicity, we further assume the initial value of ${\cal Y}_+ = {\cal Y}_-$, and so there are four possible $\CQ$-onium bound states that can contribute the diphoton excess
\begin{equation}
\bar{\mathcal Q}^{_\heartsuit} \mathcal Q^{_\heartsuit},\quad \bar
{\mathcal Q}^{_\spadesuit} \mathcal Q^{_\spadesuit},\quad \bar{\mathcal
Q}^{_\clubsuit} \mathcal Q^{_\clubsuit},\quad \bar{\mathcal
Q}^{\diamond} \mathcal Q^{\diamond}\,,
\end{equation}
where we have already combined the left and right techni-qards to form
Dirac spinors. All of them are spin-0 ($\eta_\CQ$) and have equal
masses $M_X \sim 2 m_\CQ$.

Following~\cite{Barger:1987xg,Kats:2012ym}  we describe the $SU(N_{\rm TC})$ binding potential in the Coulomb
approximation. The radial wave function $R^{(n,l)}(r)$ follows
from that of the hydrogen atom 
\begin{equation}
\left(\frac{|R^{(n,l)}(0)|^2}{M_X^3}\right)_{\rm Coul}  = \frac 1 {16
  n^3} (C_N {\bar \alpha}_{\rm TC})^3 \,,
\el{wavefunction}
\end{equation}
where $n$ is the principal quantum number, $l$ is the orbital angular
momentum, and $\bar \alpha_{\rm TC} \equiv \alpha_{\rm TC} (a_0^{-1})$
is the techni-color gauge coupling in the MS scheme, with $a_0$ the
Bohr radius of the bound state~\cite{Iwamoto:2016ral}. For fermions in the
fundamental representation, $C_N = (N_{\rm TC}^2 -1)/(2N_{\rm TC})$.
The Coulomb approximation is reliable when the non-perturbative effect
is small. In what follows, we demand that the inverse Bohr radius, which is the characteristic
scale of the bound state dynamics, is above the confinement scale:
\be \label{Coul-range}
a_0^{-1} \sim {\alpha}_{\rm TC}(a_0^{-1}) M_X > \Lambda_{\rm TC} \ee where the
confinement scale $\xL_{\rm TC}$ reads \begin{equation} \Lambda_{\rm
    TC}\sim M_X \exp\left[-\frac {6\pi}{(11 N_{\rm TC}-2N_f)
      \alpha_{\rm TC}(M_X)}\right]\,. \end{equation}  
Intuitively, (\ref{Coul-range}) states that the perturbative
treatment (like the Coulomb approximation) breaks down below the
confinement scale. The
  value of $a_0^{-1}$ ranges from 58~GeV to about 320~GeV depending on
  the value of $N_{\rm TC}$ and $Y_{\cal Q}$.

The four resonance states can be produced via photon fusion~\cite{Fichet:2015vvy,Csaki:2015vek,Anchordoqui:2015jxc,Csaki:2016raa,Harland-Lang:2016qjy,Martin:2016byf,Anchordoqui:2016rve}. 
The total photo-production cross section at LHC13 can be parametrized by
\be \sigma_{\rm LHC13} (\gamma \gamma
\to \eta_\mathcal{Q} \to \gamma \gamma) =
4.1~\text{pb}~\left(\frac{\Gamma_{\zt{total}}}{45~\text{GeV}} \right)
{\cal B}^2(\eta_{\mathcal Q} \rightarrow \gamma\gamma)
\label{shl-1}
\ee  
and the one at LHC8 by
\be \sigma_{\rm LHC8} (\gamma \gamma \to \eta_{\mathcal Q} \to \gamma \gamma) =
1.4~\text{pb}~\left(\frac{\Gamma_{\zt{total}}}{45~{\rm GeV}} \right)
{\cal B}^2(\eta_{\mathcal Q} \rightarrow \gamma\gamma)\, , \ee
showing consistency with the 95\% CL upper limit~\cite{Khachatryan:2015qba}. Actually, the ratio of
the LHC13/LHC8 partonic luminosity is largely dominated by
systematic uncertainties driven by the parton distribution functions. 
The luminosity ratio is~\cite{Harland-Lang:2016qjy}
\begin{equation}
  \frac{{\cal L}_{\gamma \gamma} (\sqrt{s} = 13~{\rm TeV})}{{\cal L}_{\gamma \gamma} (\sqrt{s} =
    8~{\rm TeV})} = 3^{+0.1}_{-0.2}, \  2.65\pm 0.15, \ 2.1 \pm0.4,
\end{equation}
for CT14QED~\cite{Schmidt:2015zda}, MRST2004~\cite{Martin:2004dh}, and
NNPDF2.3~\cite{Ball:2013hta}; respectively. We note that the predictions
of NNPDF2.3 are only marginally compatible with LHC8
data~\cite{Khachatryan:2015qba}.  

The decay width of $\eta_\CQ$ to diphotons is given
by~\cite{Iwamoto:2016ral} \be \frac{\Gamma(\eta_{\mathcal Q} \to \xg
  \xg)}{M_X} = 4 N_{\rm TC} \lb Y_{\mathcal Q}+\frac 1 2 \rb ^4 \alpha^2
\frac{|R(0)|^2}{M_X^3}\,, \ee where $R(0) \equiv R^{(1,0)}(0)$. The
total width is dominated by two major channels. One of them is to techni-gluons
$\Gamma(\eta_{\mathcal Q} \to g_{\zt{TC}} g_{\zt{TC}})$, with 
\be
\frac {\Gamma(\eta_{\mathcal Q} \to g_{\zt{TC}}
  g_{\zt{TC}})}{\Gamma(\eta_{\mathcal Q} \to
  \gamma\gamma)} = \frac{ \left(N_{\rm
      TC}^2-1\right)}{4 N_{\rm TC}^2}\frac{ \alpha_{\zt{TC}}^2} {\left(Y_{\mathcal
      Q}+\frac{1}{2}\right)^4\alpha^2} \,.  \ee
and the other is $\Gamma(\eta_{\mathcal Q} \to Z H)$ \cite{Barger:1987xg}
\be
\frac {\Gamma(\eta_{\mathcal Q} \to Z H)}{\Gamma(\eta_{\mathcal Q} \to
  \gamma\gamma)} = \frac{M_X^4} {4 {M_Z}^4}\frac{ a^2 \xa_Z^2 \left[(1-R_H-R_H)^2-4 R_H R_Z\right]^{3/2}} {4\left(Y_{\mathcal
      Q}+\frac{1}{2}\right)^4\alpha^2} \,,  \ee
where $R_i = (M_i/M_X)^2$ and $M_i$ ($i=H,Z$) are the masses of Higgs and $Z$-boson. Also $\xa_Z = \xa/(\sin^2 \xt_W \cos^2 \xt_W)$ and $a = \frac 1 2 (I_{3L} - I_{3R})$, where $I_{3L}$ ($I_{3R}$) is the isospin for the left- and right-handed techni-qards. Notice that the decay width into $ZH$ is enhanced by a factor of $\frac{M_X^4} {{M_Z}^4}$ due to the longitudinal mode of the $Z$-boson. The decay width into $HH$ would receive similar enhancement had it not been forbidden by CP symmetry ($HH$ has $J^{\zt{PC}} = 0^{++}$ while $\eta_\CQ$ has $J^{\zt{PC}} = 0^{-+}$). We will determine $\xa_{\zt{TC}}$ (as a function of $N_{\rm
  TC},Y_{\mathcal Q}$) by fitting the predicted production cross
section times branching to the observed value. Before proceeding we
note that the ATLAS excess is quite broad and probably with a large
uncertainty. The CMS excess, however, is smaller and has no clear
preference for a large width. This seems to indicate that the ATLAS
excess could be a real signal combined with a large fluctuation,
making the excess appear larger and wider than the underlying physical
signal. Throughout we assume the resonance needs to have a
signal~\cite{Kats:2016kuz}
\begin{equation}
\sigma_{\rm LHC13} (pp \to X + {\rm
  anything}) \times {\cal B} (X \to \gamma \gamma) \approx 3 -
6~{\rm fb}\, .
\label{mattS}
\end{equation}
Note that the $2$ bound states of electric charge $Y_\CQ + \frac 1 2$ contribute equally to the total cross section and dwarf the contributions from the other two states of charge $Y_\CQ - \frac 1 2$ because $\xs$ is proportional to the 8th power of the electric charge.

\begin{figure}[tpb]
\begin{minipage}[t]{0.80\textwidth}
\postscript{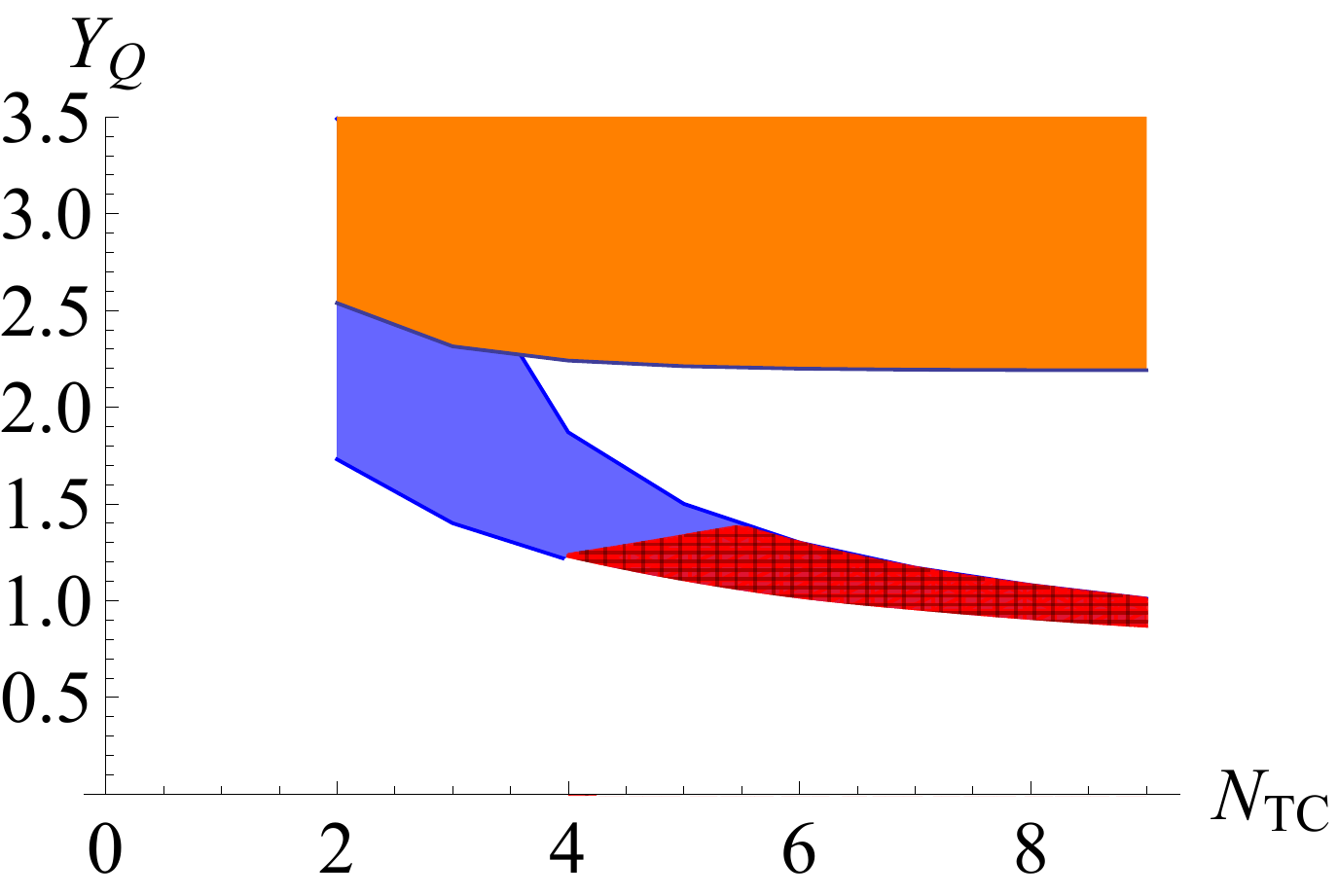}{0.78}
\end{minipage}
\begin{minipage}[t]{0.80\textwidth}
\postscript{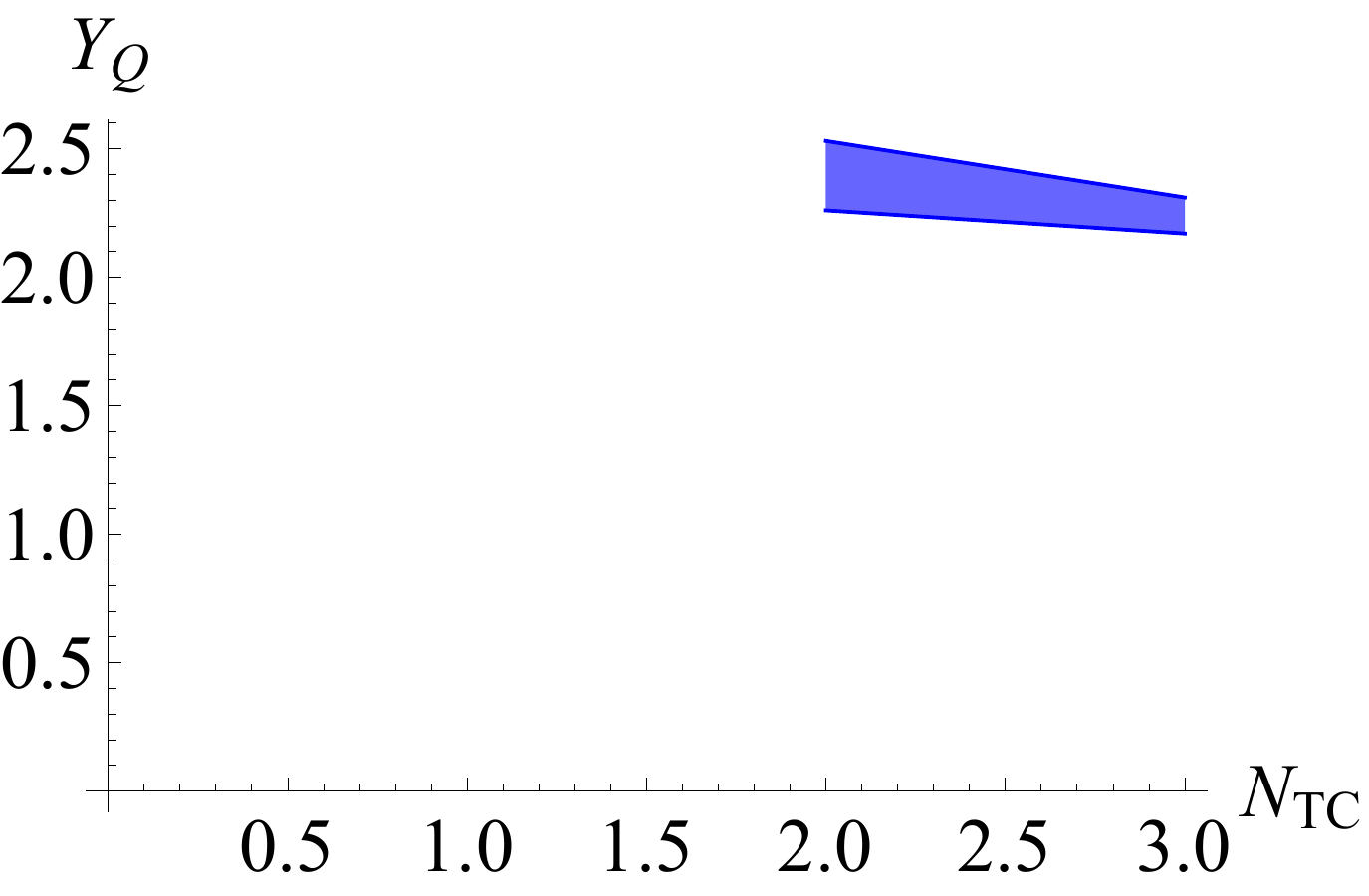}{0.78}
\end{minipage}
\caption{{\bf Top.} The (blue) banana-shape region is obtained by
demanding that the Landau pole of the Yukawa coupling is above
30~TeV and that the production cross section of $\eta_{\cal Q}$ times
its branching into diphotons is about 5~fb.  The (red) cross-hatched tail is excluded because the Coulomb approximation
fails. The (orange) band is where the contribution from decay into diphotons accounts for more than 10\% of the total width. {\bf
  Bottom.} Allowed region of the parameter space for $\Gamma(\eta_{\mathcal Q} \to Z H) < 10~\Gamma(\eta_{\mathcal Q} \to \xg
  \xg)$.}
\label{fig:1}
\end{figure}

Our results are encapsulated in Fig.~\ref{fig:1}, where we show the
region of the parameter space that can explain the diphoton signal and satisfy the bound $\Gamma(\eta_{\mathcal Q} \to Z H) < 10~\Gamma(\eta_{\mathcal Q} \to \xg \xg)$. The (blue) banana-shape region in the top panel is obtained by
demanding  both that the Landau pole of the Yukawa coupling is above
30~TeV and that the $\eta_{\cal Q}$ production cross section times
branching into diphotons (\ref{shl-1}) is 5~fb.  The (red) cross-hatched tail is excluded because the Coulomb approximation
fails. The (orange) region on the top of the figure is where the contribution to the decay width from the channel into diphotons is no longer negligible. The parameter space of the
$Y_{\cal Q} - N_{\rm TC}$ plane is further constrained by the perturbativity condition $\alpha_{\rm TC}
(a_0^{-1}) < 1$ and by the requirement $\Gamma(\eta_{\mathcal Q} \to Z
H) < 10~\Gamma(\eta_{\mathcal Q} \to \xg \xg)$. Note that within the
allowed region of the parameter space we always have $\Gamma_{\rm
  total} \leq 10~{\rm GeV}$ for each bound state. In fact, the
constraint  $\alpha_{\rm TC} (a_0^{-1})<1$ itself requires $\Gamma_{\rm total} < 11~{\rm GeV}$. The allowed region of the parameter space is significantly bounded; the hypercharge needs to lie between $2.26 \leq Y_{\mathcal Q} \leq 2.53$ for $N_{\rm TC} =2$ and be within the range $2.17 \leq Y_{\mathcal Q} \leq 2.31$ for $N_{\rm TC} = 3$. Inside this region, the partial decay widths into $W^+ W^-$ and into $Z Z$ are negligible compared to $\Gamma(\eta_{\mathcal Q} \to \xg \xg)$ and hence the corresponding bounds are satisfied trivially. 

\section{Constraints from cosmology}
\label{secIV}

Primordial nucleosynthesis provides the earliest observationally
verified landmark and constraints from big bang nucleosynthesis (BBN)
can bound the parameters of beyond SM physics models~\cite{Sarkar:1995dd}. Of particular
interest here, the techni-gluons from the decay of $\mathcal{Q}$-onium
would hadronize to techni-glueballs $G_{\zt{TC}}$ of mass $M_G \sim 7
\xL_{\rm TC}$. We must then  
verify that the dominant techni-glueball decay, $G_{\zt{TC}} \to \gamma \gamma$, 
does not drastically alter
any of the light elemental abundances synthesized during
BBN.

Note that if the $G_{\rm TC}$ decay takes place before BBN, then the photons injected
into the plasma would rapidly redistribute their energy through
scattering off background photons and through inverse
Compton scattering. The thermalization process will be particularly efficient at
plasma temperatures above 1~MeV, which is the threshold for background
$e^+e^-$ pair annihilation, and which, incidentally, coincides with time of
about 1~second. For $N_{\rm TC}=3$, the techni-glueball decay
width is given by \be \Gamma \left(G_{\zt{TC}} \to \gamma \gamma
\right) \approx \frac{(Y_\CQ + 1/2)^4 \alpha^2}{64 \pi^3}
\frac{M_{G}^3}{m_\CQ^2} \left( \frac{3 M_{G}^3}{60 m_\CQ^3} \right)^2
\, ; \ee the values for different $N_{\rm TC}$ are of the same
order~\cite{Chen:2005mg,Juknevich:2009ji,Juknevich:2009gg}. A straightforward substitution shows that in our model $G_{\rm
  TC}$ 
decay would not not alter BBN as the techni-glueball lifetime $<
10^{-28}~{\rm s}$.

\section{Looking ahead}
\label{secV}

In this work we have attempted to associate the possible event excess
in the diphoton invariant mass spectrum around 750~GeV, as indicated
in LHC13 data, with bound states of a new asymptotically free gauge
theory.  In particular, we have constructed a minimal model with one
generation of uncolored fermions $\mathcal Q$, in which the $\mathcal
Q$ fields transform as $SU(2)$ doublets and singlets, and are
invariant under the $SU(N_{\rm TC})$ transformation of a hidden strong
gauge interaction to be explored during the LHC Run II data taking
period. Since the new colorless fields transform as $SU(2)$ doublets
they can acquire a mass through Yukawa interactions with the
electroweak Higgs doublet, as quarks and leptons. We parametrized the
${\mathcal Q}$-onium bound state using the Coulomb approximation and
gave a numerical fit to the diphoton excess consistent with no strong
coupling effects up to about 30~TeV. We have shown that allowed
hypercharges lie within the range of $2.26 \leq Y_{\mathcal Q} \leq
2.53$ for $N_{\rm TC} =2$ and $2.17 \leq Y_{\mathcal Q} \leq 2.31$ for
$N_{\rm TC} = 3$.  The smoking gun for the model would be the
discovery of $\eta_\mathcal{Q} \to ZH$.

In closing, we note that the $\mathcal{Q}$ Yukawa couplings drive the
quartic Higgs coupling to negative values in the ultraviolet and the
SM scalar effective potential develops an instability above about
100~TeV. As noted elsewhere~\cite{Anchordoqui:2012fq} (see also~\cite{Kadastik:2011aa,EliasMiro:2012ay}) the potential instability of
the electroweak vacuum can be evaded if the scalar sector contains a
hidden heavy scalar singlet (with a large vacuum expectation value),
which mixes with the SM Higgs doublet.  The quartic interaction
between the heavy scalar singlet and the Higgs doublet leads to a
positive tree-level threshold correction for the Higgs quartic
coupling, which is very effective in stabilizing the potential. In
addition, the hidden scalar singlet could deliver mass terms for the
vector-like hidden quarks $Q$ and/or ${\mathpzc Q}$~\cite{inprepa}.

\acknowledgments{We would like to thank Xerxes Tata, Vernon Barger,
  and Chaehyun Yu for valuable discussions. L.A.A.  is supported by
  U.S. National Science Foundation (NSF) CAREER Award PHY1053663 and
  by the National Aeronautics and Space Administration (NASA) Grant
  No. NNX13AH52G; he thanks the Center for Cosmology and Particle
  Physics at New York University for its hospitality.  X.H.  is
  supported by the MOST Grant 103-2811-M-003-024; he thanks the
  Institute of Modern Physics at Northwest University for its
  hospitality.  Any opinions, findings, and conclusions or
  recommendations expressed in this material are those of the authors
  and do not necessarily reflect the views of the National Science
  Foundation.}

\end{document}